# Analytic evaluation of diffuse fluence error in multi-layer scattering media with discontinuous refractive index

**Adrian C Selden*[‡]**

20 Wessex Close, Faringdon, Oxfordshire, SN7 7YY, United Kingdom

*email: adrian_selden@yahoo.com

**ABSTRACT**

A simple analytic method of estimating the error involved in using an approximate boundary condition for diffuse radiation in two adjoining scattering media with differing refractive index is presented. The method is based on asymptotic planar fluences and enables the relative error to be readily evaluated without recourse to Monte Carlo simulation. Three examples of its application are considered: (1) evaluating the error in calculating the diffuse fluences at a boundary between two media with differing refractive index and dissimilar scattering properties (2) the dependence of the relative error in a multi-layer medium with discontinuous refractive index on the ratio of the reduced scattering coefficient to the absorption coefficient $\mu_s'/\mu_a$ (3) the parametric dependence of the error in the radiant flux $J_s$ at the surface of a three-layer medium. The error is significant for strongly forward biased scattering media with non-negligible absorption and is cumulative in multi-layered media with refractive index increments between layers.

# 1. INTRODUCTION

The transport of light in strongly scattering turbid media, such as biological tissue, is generally modelled as a diffusion process, described by the diffusion equation for the fluence $\varphi$ [1, 2]. In condensed scattering media, $\varphi$ is dependent on the refractive index n, such that $\varphi/n^2$ is conserved, enabling modelling of the diffusion of light in media with spatially varying refractive index [3]. However, a finite discontinuity in refractive index, as occurs at the boundary between two scattering media of differing refractive index, gives rise to Fresnel reflection, such that $\varphi/n^2$ is discontinuous, with a discontinuity $\Delta\varphi$ proportional to the diffuse radiant flux J at the boundary [4-7]

$$\Delta\varphi = (n_2/n_1)^2\varphi_1 - \varphi_2 = C(n_2/n_1)J \qquad (1)$$

where $C(n_2/n_1)$ is a smoothly varying function of the index ratio $n_2/n_1$ tabulated in [4].

The discontinuity in $\varphi/n^2$ depends on the ratio of refractive indices at the boundary, and is small for modest index ratios. Monte Carlo simulation of diffuse light transport across boundaries between turbid media with different refractive indices has shown that the error introduced when this correction is ignored is generally less than 10% for weakly absorbing scattering media e.g. biological tissue illuminated with infra-red radiation [6]. An analytical solution for time-dependent diffusion between adjacent half-spaces presented in [8] supports this conclusion. However, the error is cumulative in multi-layer scattering media, and increases significantly for strongly forward-biased scattering and non-negligible absorption. A simple analytic method of assessing the error incurred in these circumstances when the discontinuity in $\varphi/n^2$ is not taken into account would therefore be useful, and is discussed below. Errors in modelling the diffuse fluence $\varphi$ in turbid media can lead to systematic errors in diffuse transmittance and reflectance. They also introduce errors in scattering and absorption coefficients inferred from reflectance measurements and thus to errors in quantitative image reconstruction via diffuse optical tomography. Errors in estimating internal diffuse fluence may also impact on photodynamic therapy (PDT).

## 2. THEORY

### 2.1. INTERFACE ERROR

With the definitions [9]

$$\varphi = \tfrac{1}{2} \int I(\xi)d\xi \qquad \xi \in |-1, 1| \qquad (2a)$$

$$J = \tfrac{1}{2} \int I(\xi)\xi d\xi \qquad \xi \in |-1, 1| \qquad (2b)$$

where $I(\xi)$ is the angular intensity distribution and $\xi$ the direction cosine with respect to the positive z-axis, we find the mean cosine of the radiant intensity distribution $\langle\xi\rangle = \int I(\xi)\xi d\xi / \int I(\xi)d\xi = J/\varphi$. Dividing eqn (1) by $\varphi_1$, we find the relative error

$$\Delta\varphi/\varphi_1 = (n_2/n_1)^2 - \varphi_2/\varphi_1 = C(n_2/n_1)J_1/\varphi_1 = C(n_2/n_1)\langle\xi_1\rangle \qquad (3a)$$

Similarly
$$\Delta\varphi/\varphi_2 = (n_2/n_1)^2\varphi_1/\varphi_2 - 1 = C(n_2/n_1)J_2/\varphi_2 = C(n_2/n_1)\langle\xi_2\rangle \qquad (3b)$$

Thus the error in applying the approximate boundary condition is directly proportional to the mean cosine $\langle\xi\rangle$ of the angular intensity distribution at the boundary. This is illustrated in Fig 1, where the ratio $J/\varphi$ vs. refractive index ratio n for a given relative error $\Delta\varphi/\varphi$ (left hand axis) and the relative error $\Delta\varphi/\varphi$ vs. n for a given ratio $J/\varphi$ (right hand axis) are plotted. Thus the larger the index ratio n, the smaller the ratio $J/\varphi$ required for a given relative error $\Delta\varphi/\varphi$ and the larger the relative error $\Delta\varphi/\varphi$ for a given ratio $J/\varphi$ i.e. for a given angular radiance distribution $I(\xi)$ with mean cosine $\langle\xi\rangle$.

The discontinuity in diffuse fluence $\varphi$ quantified by eqn (1) also implies a discontinuity in the mean cosine $\langle\xi\rangle$ of the radiance distribution $I(\xi)$ at the boundary viz. $\langle\xi_1\rangle \neq \langle\xi_2\rangle$. To find the magnitude of the error in $\varphi$ for a specific case requires numerical evaluation of the boundary fluxes [6]. However, an estimate (lower bound) can be made in terms of the mean cosine of the asymptotic angular radiance $\langle\xi\rangle_{as}$ given by [10]

$$\langle\xi\rangle_{as} = (1-a)/\mu_{as} \qquad (4)$$

for scattering albedo $a = \mu_s/(\mu_s+\mu_a) = \mu_s/\mu_e$, where $\mu_s$ is the scattering coefficient, $\mu_a$ the absorption coefficient, $\mu_e = \mu_s+\mu_a$ the extinction coefficient and $\mu_{as}$ the asymptotic attenuation coefficient. In the $\delta$-P1 approximation [2], $\mu_{as}$ is replaced by $\mu_{eff}$, the effective attenuation coefficient. $\mu_{eff}$ and the mean cosine $<\xi'>_{as}$ are determined by the reduced scattering coefficient $\mu'_s$ and the absorption coefficient $\mu_a$ viz.

$$\mu_{eff} = [\mu_a/D']^{1/2} = [3\mu_a(\mu_a + \mu'_s)]^{1/2} = (3\mu_a\mu_{tr})^{1/2} \qquad (5a)$$

$$<\xi'>_{as} = \mu_{eff}D' = [\mu_aD]^{1/2} = [\mu_a/3(\mu_a + \mu'_s)]^{1/2} = (\mu_a/3\mu_{tr})^{1/2} \qquad (5b)$$

where $\mu'_s = \mu_s(1-g)$, g is the scattering asymmetry, the transport coefficient $\mu_{tr} = \mu_a + \mu'_s$ and $D = 1/3\mu_{tr}$ is the diffusion coefficient [1]. Thus $\mu_{eff} \Rightarrow 0$, $<\xi'>_{as} \Rightarrow 0$ (isotropic radiance) when $\mu_a \Rightarrow 0$ (zero absorption). More precise evaluation of the effective attenuation coefficient $\mu_{eff}$ (and of $<\xi'>_{as}$), required for forward-biased scattering in absorbing media, involves higher moments of the phase function [11]. In the δ-P3 approximation, the effective attenuation coefficient

$$\mu_{eff}^{(3)} = [(\beta - (\beta^2 - 36\gamma)^{1/2})/18]^{1/2} \qquad (6a)$$

where $\beta = 27\mu_a\mu'^{(1)}_t + 28\mu_a\mu'^{(3)}_t + 35\mu'^{(2)}_t\mu'^{(3)}_t$, $\gamma = 105\mu_a\mu'^{(1)}_t\mu'^{(2)}_t\mu'^{(3)}_t$, $\mu'^{(m)}_t = \mu_a + \mu_s(1-g^m)$, m = 1, 2, 3 [12, 13] and the mean cosine of the asymptotic radiance

$$<\xi'>_{as}^{(3)} = (1 - a')/\mu_{eff}^{(3)} \qquad (6b)$$

The dependence of the relative error $\Delta\varphi/\varphi$ on scattering asymmetry g is shown in Fig 2 for scattering albedoes in the range $a \in |0, 0.99|$ for accurate values of $\mu_{eff}$ [11]. It can be seen that $\Delta\varphi/\varphi$ is only weakly dependent on scattering asymmetry for g<0 (backward-biased scattering), even for strong absorption ($a = 0.2$ i.e. $\mu_a = 4\mu_s$), while increasing rapidly for forward-biased scattering (g>0), approaching 10% for g≥0.99 when $a$=0.9. In the δ-P1 approximation, the scattering asymmetry is reduced: g'$\in |0, 0.5|$ for g$\in |0, 1|$, but so is the scattering albedo: $a' = \mu_s'/(\mu_s'+\mu_a)$ via the reduced scattering coefficient $\mu_s' = \mu_s(1-g)$, potentially offsetting a reduced error in $\varphi$. The error in diffuse fluence increases for interfaces with higher index ratios, $\Delta\varphi/\varphi$ exceeding 20% for g = 0.95 when n = 1.25.

## 2.2 DIFFUSION EQUATION

The diffuse fluence $\varphi$ obeys the steady-state diffusion equation [1]

$$\partial^2\varphi/\partial z^2 + S(z)/D = \mu_{eff}^2\varphi \qquad (7)$$

where S(z) is the source distribution, D the diffusion coefficient and $\mu_{eff}$ the effective attenuation coefficient. The radiant flux (nett energy flow) J is given by Fick's law [1]

$$J = -D\partial\varphi/\partial z \qquad (8)$$

Solutions of the diffusion equation eq (7), subject to the boundary conditions, define the distribution of the diffuse fluence $\varphi$ and radiant flux J in scattering media with discontinuous refractive index.

## 2.3. BOUNDARY CONDITIONS

The boundary conditions at an interface between two diffusive scattering media with differing refractive indices $n_1$, $n_2$ are

$$(n_2/n_1)^2\varphi_1 - \varphi_2 = C(n_2/n_1)J \qquad (9a)$$

for diffuse fluence $\varphi$, where $n = n_2/n_1 > 1$ and $C(n_2/n_1) \propto (n_2/n_1 - 1)^{3/2}$ for $n_2/n_1 - 1 \ll 1$ [4]

and

$$-D_1\partial\varphi_1/\partial z = -D_2\partial\varphi_2/\partial z \qquad (9b)$$

for conservation of radiant flux $J = -D\partial\varphi/\partial z$ across the interface. These are applied to specific cases in Sec. 3 below. The error $\Delta\varphi$ in diffuse fluence resulting from application of the approximate boundary condition [14]

$$(n_2/n_1)^2\varphi_1 - \varphi_2 \approx 0 \qquad (10)$$

is proportional to J (eq (1)).

## 2.4. DIFFUSE FLUENCE EQUATIONS

To proceed further, we require solutions of the diffusion equation for two adjoining layers satisfying the boundary conditions eqns (9a, 9b). To simplify the analysis, we consider planar asymptotic solutions for $\varphi_1$ and $\varphi_2$ in the respective scattering media [15]

$$\varphi_1(z) = a_1\exp(\mu_{eff-1}z) + b_1\exp(-\mu_{eff-1}z) \quad z<0 \quad (11a)$$

$$\varphi_2(z) = a_2\exp(\mu_{eff-2}z) + b_2\exp(-\mu_{eff-2}z) \quad z>0 \quad (11b)$$

with effective attenuation coefficients $\mu_{eff-1}$, $\mu_{eff-2}$; the z-axis is taken perpendicular to the interface at $z = 0$. Inserting these solutions in eqns (9a, 9b), we find

$$\varphi_1(0) = 2K/(1+K) \quad (12a)$$

$$\varphi_2(0) = 2D_1\mu_{eff\text{-}1}/[D_2\mu_{eff\text{-}2}(1+K)] \quad (12b)$$

where
$$K = [1+ C(n_2/n_1)D_2\mu_{eff\text{-}2}]K' \quad n_2>n_1 \quad (12c)$$

$$K = [1+ C(n_1/n_2)D_2\mu_{eff\text{-}2}/(n_1/n_2)^2]K' \quad n_2<n_1 \quad (12c')$$

and
$$K' = [D_1\mu_{eff\text{-}1}/(n_2/n_1)^2 D_2\mu_{eff\text{-}2}] \quad (12d)$$

assuming a semi-infinite medium (half-space) for $z>0$ i.e. $a_2 = 0$ for $\varphi_2(z)\Rightarrow 0$ as $z\Rightarrow\infty$.

Eqns (12a,b,c,c',d) enable comparison of the diffuse boundary fluences $\varphi_1(0)$, $\varphi_2(0)$ with those satisfying the approximate boundary condition eqn (10), which follow on setting $C(n_2/n_1) = 0$ in eq (12c) i.e. for $K \Rightarrow K'$. Analytic evaluation of the fractional flux error in terms of the refractive index ratio $n=n_2/n_1$ and the diffusion parameters $D_1\mu_{eff\text{-}1}$, $D_2\mu_{eff\text{-}2}$ via the scattering asymmetry g and scattering albedo *a* can then be made. Accurate values of $D_1\mu_{eff\text{-}1}$ and $D_2\mu_{eff\text{-}2}$ for forward-biased anisotropic scattering in absorbing media (*a*<1) may be calculated from the phase function $p(\xi)$ and scattering albedo *a* [10]. Alternatively, the mean cosine $<\xi>_{as}$ of the asymptotic radiance can be obtained from eq (4) and used in place of $D\mu_{eff}$. Only $\mu_{eff}$ need be calculated in this case, either analytically in the P1 or P3 approximations [12, 13], or numerically for higher accuracy [11].

## 3. RESULTS

### 3.1 INTERFACE

The boundary condition eq (9a) has been evaluated analytically by Shendeleva for time dependent diffusion in adjoining media with isotropic scattering, and validated by Monte Carlo simulation [3]. Validation of the analytic method presented here is provided by comparison with the results of Ripoll and Nieto-Vesperinas, who evaluated the error using numerical methods [6]. Fig. 3 shows the relative errors in the diffuse fluences at the common boundary between two adjoining media vs. the refractive index $n_2$ of a scattering medium adjoining an aqueous scattering medium ($n_1$ = 1.333), as calculated from the analytic formulae (eqs. 12a,b,c,c',d) above. The results show precise agreement with the numerical data (plotted points) of [6], confirming the validity of the simpler analytic method, which can therefore replace the previous numerical methods for rapid evaluation of the error in similar cases.

Fig. 4 shows the dependence of the fractional errors $\Delta\varphi/\varphi$ in the diffuse fluences vs. scattering albedo $a$ for two adjoining media with disparate scattering parameters (g = 0, 0.95), calculated in the P1 and P3 approximations to the diffusion parameters for Henyey-Greenstein scattering [16], with refractive index ratio n = 1.41/1.34 = 1.06 (tissue/aq). Initially, the error increases rapidly with absorption ($a$<1) for g = 0, reaching a broad maximum $\Delta\varphi/\varphi$ ~ 6% when $a$ ~ 0.6; in contrast, the error for g = 0.95 in the adjoining medium increases quasi-linearly to ~ 5% when $a$ = 0. The results show that the P1 approximation seriously underestimates the error in diffuse fluence (by ~40%), while P3 is ≤10% low compared with the accurate value, and is preferred for analytic evaluation of $\Delta\varphi/\varphi$. Overall, the error increases sharply when there is non-negligible absorption in a turbid medium with strongly forward biased scattering

### 3.2. MULTI-LAYERS

More generally, for diffusion of light in multiple layers of finite thickness, the diffuse fluence $\varphi_k(z)$ in the kth layer may be expressed as [15]

$$\varphi_k(z) = a_k \exp(\mu_{eff\text{-}k} z) + b_k \exp(-\mu_{eff\text{-}k} z) \qquad (13)$$

with a similar expression for $\varphi_{k+1}(z)$ in the (k+1)th layer. The boundary conditions [4, 7]

$$(n_{k+1}/n_k)^2 \varphi_k(z_k) - \varphi_{k+1}(z_k) = C(n_{k+1}/n_k) J_k(z_k) \qquad (14a)$$

$$- D_k \nabla \varphi_k(z_k) = - D_{k+1} \nabla \varphi_{k+1}(z_k) \qquad (14b)$$

yield the simple recurrence relations (for $n = n_{k+1}/n_k > 1$)

$$a_{k+1} = \tfrac{1}{2}\{[n^2+1+C(n)D\mu_{eff}]a_k + [n^2-1-C(n)D\mu_{eff}]b_k \exp(-2\mu_{eff} h)\} \qquad (15a)$$

$$b_{k+1} = \tfrac{1}{2}\{[n^2-1+C(n)D\mu_{eff}]a_k \exp(2\mu_{eff} h) + [n^2+1-C(n)D\mu_{eff}]b_k\} \qquad (15b)$$

when $D_{k+1}\mu_{eff\text{-}k+1} = D_k\mu_{eff\text{-}k} = D\mu_{eff}$ and $\mu_{eff\text{-}k+1} h_{k+1} = \mu_{eff\text{-}k} h_k = \mu_{eff} h$, where $h_k$, $h_{k+1}$ are the widths of the kth and (k+1)th layers, enabling the coefficients $a_{k+1}$, $b_{k+1}$ to be related $a_k$, $b_k$. The results for the approximate boundary condition eqn (10) are obtained on setting C(n) = 0 in eqns (15a, b). Successive application of these relations yields the coefficients $a_k$, $b_k$ for all the layers involved, with appropriate boundary conditions chosen for the first and last [15]. A parallel set of coefficients $a_k'$, $b_k'$, for $C(n) = 0$, enables direct comparison of the accurate and approximate fluxes $\varphi_k$, $\varphi_k'$ in each layer, and thus evaluation of the cumulative error for the multi-layer system. This is illustrated in Fig 5, with $\Delta\varphi/\varphi = 1.4\%$ at a single interface (for $a = 0.995$, $g = 0.95$, $n = 1.1$), the cumulative error increasing with the total number of layers, exceeding 30% for 5 layers when $\mu_s'/\mu_a = 1$. For multi-layer media with higher index ratios, or for a larger number of layers, the cumulative error can easily exceed 100%.

### 3.3. PERTURBING LAYER

The dependence of the diffuse reflectance of a layered medium on changes in the optical properties of a sub-surface layer is of special interest [17]. We consider a simple three-layer model comprising two plane parallel layers supported on a semi-infinite layer (half-space) and vary the properties of the middle layer to illustrate this. The arrangement is sketched in Fig 6. The optical properties are given in Table I. The problem is analysed via the equations for the diffuse fluences in the three regions:

Half-space $\quad\quad\quad \varphi_3(z) = b_3 \exp(-\mu_{eff-3}\, z)$ (16a)

Mid-layer $\quad\quad\quad \varphi_2(z) = a_2 \exp(\mu_{eff-2}\, z) + b_2 \exp(-\mu_{eff-2}\, z)$ (16b)

Surface layer $\quad\quad \varphi_1(z) = a_1 \exp(\mu_{eff-1}\, z) + b_1 \exp(-\mu_{eff-1}\, z) - S_0 \exp(-\mu_{tr} z)$ (16c)

the diffuse fluence $\varphi_3(z)$ decaying exponentially in the half-space, the flux $\varphi_2(z)$ in the mid-layer having both exponential terms and the flux $\varphi_1(z)$ in the surface layer including the exponential source $S_0 \exp(-\mu_{tr} z)$ for plane parallel illumination at the surface [2]. Boundary conditions defined in eqns (14a, b) are applied successively at the interfaces to evaluate the coefficients $a_k$, $b_k$ for $k = 1, 2, 3$. The flux $\varphi_1(z)$ in the surface layer is extrapolated to zero a finite distance beyond the surface equal to the linear extrapolation distance $z_b$ determined by the refractive index $n_1$ [4]. The radiant flux at the surface $J_s = -D_1 d\varphi_1/dz|_s = -D_1 \varphi_1(0)/z_b$. A set of coefficients $a_k'$, $b_k'$ for $C(n_{k+1}/n_k) = 0$ at each interface yields the uncorrected surface flux $J_s'$, allowing the relative error $\Delta J_s/J_s$ (and hence the relative error in diffuse reflectance) to be determined. A representative set of curves showing the dependence of $\Delta J_s/J_s$ on the ratio $\mu_s'/\mu_a$ for selected values of the refractive index ratio n = $n_{21}$ is plotted in Fig 7, the relative error in surface flux $\Delta J_s/J_s$ increasing with n and $\mu_a$, reaching 30% for n = 1.5 and $\mu_s'/\mu_a$ = 1 (*a'* = 0.5) i.e. for $\mu_s/\mu_a$ = 20 (*a* = 0.95, g = 0.95). In general, the error is larger (smaller) for higher (lower) values of scattering asymmetry g and also increases with absorption.

## 4. DISCUSSION

The analytic method of asymptotic planar fluxes enables straightforward evaluation of the error in diffuse fluence φ without recourse to Monte Carlo simulation. The magnitude of the error is readily found from the optical properties of the adjoining scattering media, namely the refractive index ratio, the reduced scattering coefficients, the absorption coefficients and scattering asymmetries. It is simply expressed via the product C(n)<ξ>, where C(n) is a monotonically increasing function of the index ratio n = $n_2/n_1$ [4] and <ξ> is the mean cosine of the boundary radiance, a key result of the above analysis. This is approximated by the asymptotic mean cosine <ξ>$_{as}$ of the radiance far from the boundary, expressed in terms of scattering albedo *a* and diffuse attenuation coefficient $\mu_{eff}$. An equivalent formula for the error is C(n)D$\mu_{eff}$, where D is the diffusion coefficient. The dependence of flux error on scattering asymmetry g is of some interest viz. for forward-biased scattering in turbid media with near negligible absorption (typical of biological tissue in the near IR [6]); the error increases rapidly as g⇒1, but is virtually independent of scattering asymmetry for negative values g<0 (Fig 2). For forward-biased scattering media with non-negligible absorption e.g. biological media in the visible spectrum, the error becomes progressively less dependent on scattering asymmetry as absorption increases, ultimately becoming independent of g in the limit of zero scattering albedo *a*⇒0.

Having found a simple means of estimating the magnitude of the flux error at a boundary, accurate formulae for the diffuse fluence φ in adjoining media incorporating the correction term C(n)D$\mu_{eff}$ are obtained from the boundary condition eq (1). It is noted that the correction applies to the diffuse fluence distribution throughout the turbid medium, not simply at the boundaries. In the case of multi-layer media, repeated application of the boundary conditions yields recurrence relations for the coefficients $a_k$, $b_k$ of the formula for the diffuse fluence. A corresponding set of coefficients $a_k'$, $b_k'$ obtained from the approximate boundary condition (with $C(n) = 0$) enables the flux error in successive layers to be obtained and its parametric dependence on the optical properties of the layers to be investigated. This is of importance for comparison with experimental determination of diffuse reflectance, and also for the inverse problem of determining optical constants from reflectance measurements.

The principal aim of the present work was to provide a simple analytic method for estimating the error incurred in using the approximate form of the boundary condition (eq 10). This has been checked against the computational results of [6] (Sec. 3.1 and Fig 3) and illustrated with several examples relevant to biomedical optics. The work presented here concerns analysis of the error involved in applying the approximate diffusion boundary condition (eq 10), rather than the error in using diffusion theory per se. Thus a numerical evaluation would merely quantify the 'error within the error', whereas the analytic method provides a simple means of estimating its magnitude. The analytic approach was never intended to replace accurate radiative transfer computations where these are merited e.g. Phillips and Jacques [17], but rather as a simple check on the diffusion approximation e.g. the widely used $\delta$-P1 formulation [2].

## 5. CONCLUSION

A simple analytic method of estimating the error involved in applying a commonly used approximate boundary condition for diffuse radiation in two adjoining scattering media with differing refractive index has been presented. The method is based on asymptotic planar fluxes and enables the relative error to be readily evaluated without recourse to Monte Carlo simulation. Three examples of its application have been considered: (1) evaluating the error in calculating the diffuse fluences at a boundary between two media with differing refractive index and dissimilar scattering properties (2) the dependence of the relative error in diffuse fluence $\varphi$ in a multi-layer medium with discontinuous refractive index on the ratio of the reduced scattering coefficient to the absorption coefficient $\mu_s'/\mu_a$ (3) the dependence of the relative error in radiant flux $J_s$ at the surface of a three-layer medium on $\mu_s'/\mu_a$. In addition to its dependence on refractive index ratio n = $n_2/n_1$ via the function C(n), the fluence error increases with scattering asymmetry g in forward biased scattering media, and is cumulative in multi-layered media with refractive index increments between layers. The same methodology can be applied to multi-layer problems with cylindrical symmetry, the system being first converted to planar geometry via a Hankel transform, to allow 1D analysis (as here), followed by reconversion of the solution to cylindrical symmetry via an inverse Hankel transform [15].

## Table I

## Optical properties of the three layer system

| Layer | $\mu'_s$ mm$^{-1}$ | $\mu_a$ mm$^{-1}$ | d mm | n |
|---|---|---|---|---|
| 1 | 1 | 0.001 | 3 | 1.0 |
| 2 | 1 | 0.001-1 | 3 | 1.1-1.5 |
| 3 | 1 | 0.001 | inf. | 1.0 |

**FIGURE CAPTIONS**

Fig. 1. The ratio J/φ vs. refractive index ratio n for a given relative error in diffuse fluence Δφ/φ (left hand axis) and the relative error Δφ/φ vs. n for a given ratio J/φ (right hand axis)

Fig. 2. Relative error in diffuse fluence Δφ/φ vs scattering asymmetry g for a range of scattering albedoes *a*, the error increasing rapidly for g>0.9. The limiting case for zero scattering (*a*=0) is indicated by the horizontal dashed line; the vertical dotted line marks the maximum asymmetry in the δ–P1 approximation (g' = 0.5). Index ratio n = 1.06.

Fig. 3 Comparison of analytic results for diffuse fluence error Δφ/φ (curves) with numerical results (data points from Fig 8 in [5]) at the interface between an aqueous scattering medium ($n_1$ = 1.333) and a scattering medium with refractive index $n_2$ varied in the range 1≤ $n_2$ ≤2. Scattering parameters: $\mu_{s1}'$ = 15 cm$^{-1}$, $\mu_{a1}$ = 0.035 cm$^{-1}$, $\mu_{s2}'$ = 10 cm$^{-1}$, $\mu_{a2}$ = 0.24 cm$^{-1}$, g = 0.8.

Fig. 4. Relative error in diffuse fluence Δφ/φ vs scattering albedo *a* on either side of the interface between two homogeneous media with disparate scattering parameters: g = 0 (upper curves), g = 0.95 (lower curves) for index ratio n = 1.06. The filled squares (■) are data points calculated with accurate values of the diffusion parameters D, κ [10, 11].

Fig. 5. Relative error in diffuse fluence Δφ/φ vs. $\mu_s'/\mu_a$ in layer 1 (upper points) and layer 5 (lower points) of a 5-layer medium on a half-space (g = 0.95, index ratios n = 1.1); open squares (□) P1 (diffusion) values, filled squares (■) accurate values of diffusion parameters D, κ.

Fig. 6. Schematic of three-layer system, comprising two finite layers on a semi-infinite substrate, subject to plane illumination normally incident on the first layer

Fig. 7. Relative error in diffuse radiant flux $\Delta J_s/J_s$ at the surface of a three-layer system (two finite layers on a semi-infinite substrate) vs. the ratio $\mu_s'/\mu_a$ in the middle layer (layer 2) with refractive index in the range n = 1.1- 1.5 (Table I).

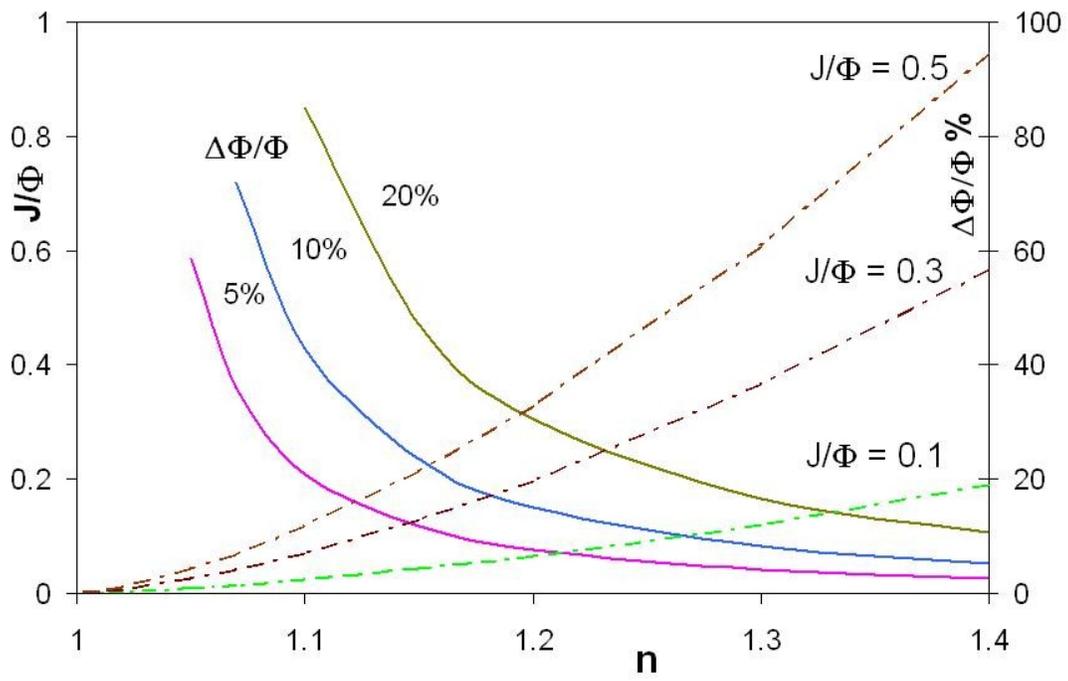

Fig 1

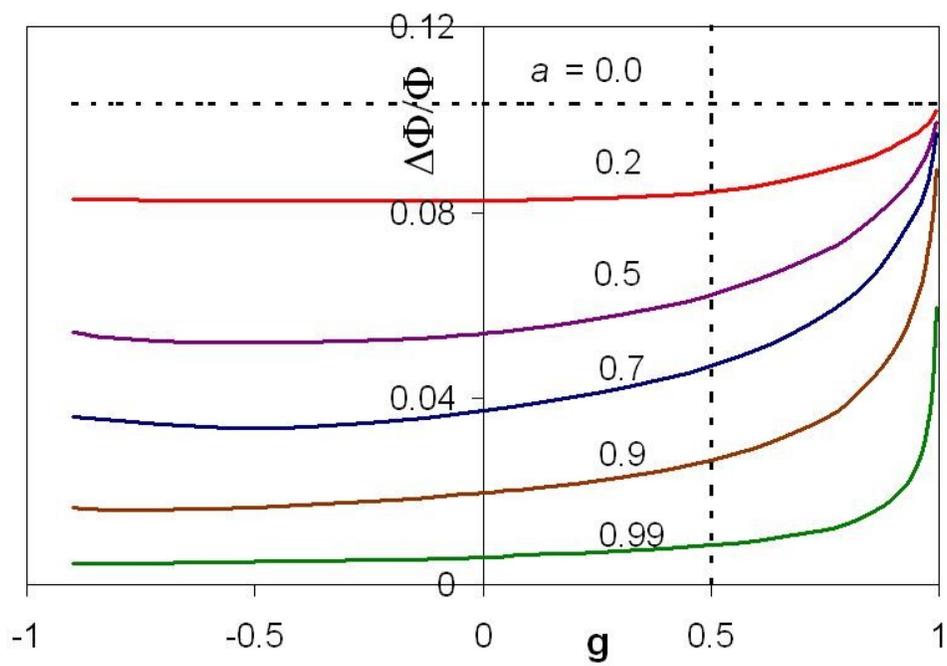

Fig 2

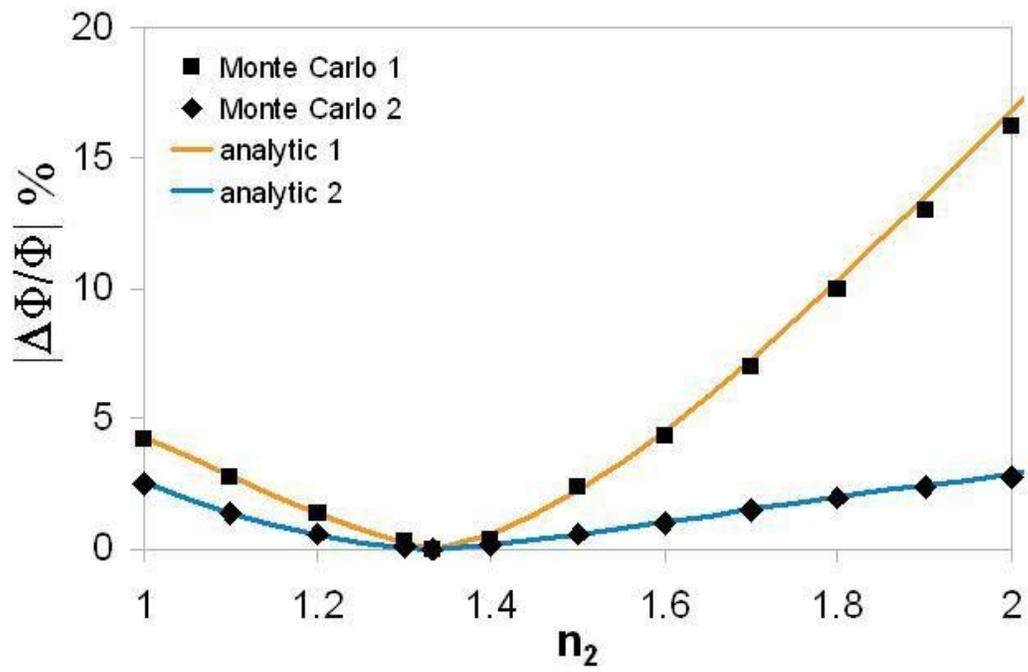

Fig 3

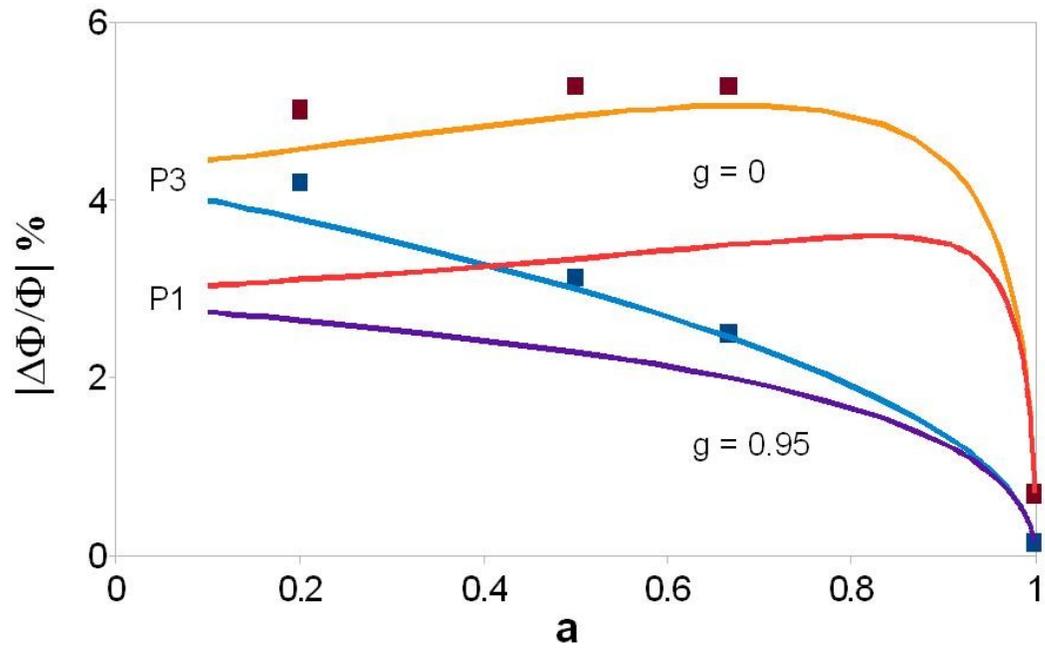

Fig 4

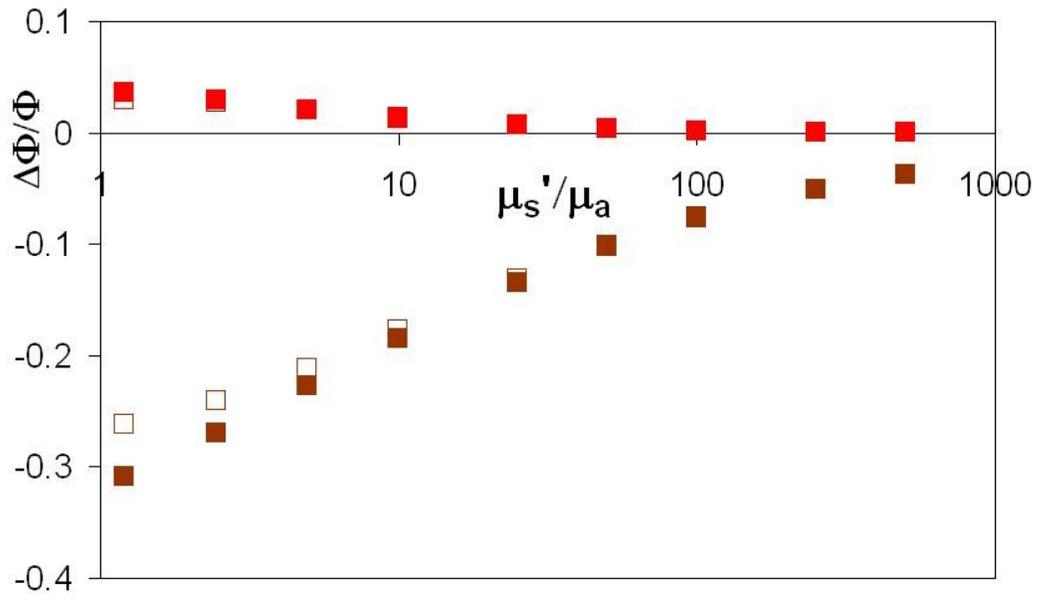

Fig 5

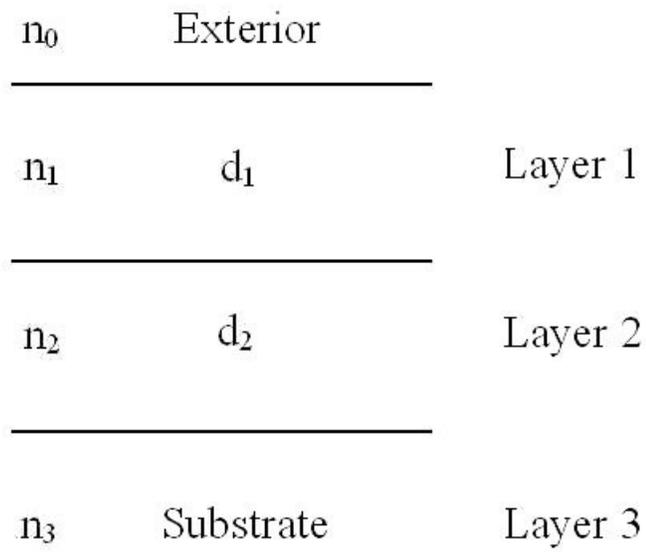

Fig 6

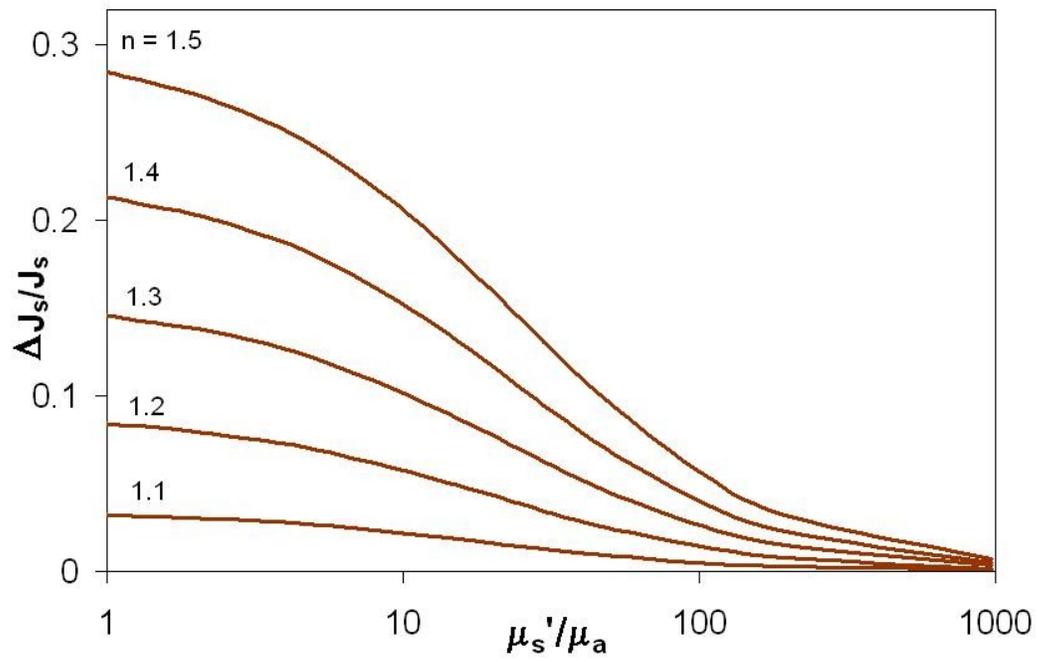

Fig 7